\documentclass[prb,twocolumn,superscriptaddress,showpacs]{revtex4}
\usepackage{graphicx}
\usepackage{amssymb,amsmath}
\begin{document}

\title{Partial localization of correlated electrons: spin dependent masses,\newline saturated  ferromagnetism, and effective s-d model.}

\author{Jakub J\c{e}drak}
\author{ Jozef Spa\l ek}
\affiliation{Instytut Fizyki im. Mariana Smoluchowskiego, Uniwersytet Jagiello\'{n}ski, Reymonta 4, PL-30059 Krak\'ow, Poland}
\author{Gertrud Zwicknagl}
\affiliation{Institut f\"{u}r Matematische Physik, Technische Universit\"{a}t Braunschweig, Mendelssohnstrasse 3, 38106 
Braunschweig, Germany}

\date{\today}

\begin{abstract}
We determine the localization threshold in a partially filled and orbitally degenerate model of correlated electrons. Particular emphasis is put on a non-integer band filling,  when the system decomposes into the localized and the itinerant subsystems; this situation is described by an effective s-d model. A simultaneous transition to the ferromagnetic state is discussed as driven by the Hund's rule coupling. Dependence of the quasiparticle mass on the spin direction appears naturally in the ferromagnetic phase and is attributed to the electron correlation effects, as is also a metamagnetic transition in an applied field.
Although the main results have been obtained within the saddle point slave-boson approach, their qualitative features are discussed in general terms, i.e. as a transition from quantum-mechanical indistinguishability of particles to the two-component situation. A comparison with the situation for the orbitally nondegenerate band is also briefly mentioned.
\end{abstract}

\maketitle

\section{\label{sec:1} Introduction}
The longstanding problem\cite{Hubbard} of the  dual (\textit{localized-itinerant}) nature of correlated electrons has been addressed recently  in the context of heavy-fermion systems containing $5f$ states due to U ions and termed \textit{the partial localization}.  \cite{gz0} It appears only when the occupancy $n_{f}$ of the $5f$ states is a noninteger number and exceeds unity, as is expected for UPt$_{3}$, UPd$_{2}$Al$_{3}$ or URu$_{2}$Si$_{2}$, even though that a relatively strong hybridization of $5f$ states with $6d$-$7s$ states takes place. Therefore, such a decomposition of the quantum-mechanically indistinguishable electrons into two separate subsystems must be accompanied by a phase transition and attributed to the correlation effects such as the Hund's rule exchange interaction, particularly when combined with the direct Coulomb (Hubbard) interaction. Such behavior should be absent for the corresponding cerium intermetallic compounds (such as CeCu$_{2}$Si$_{2}$, CeAl$_{3}$ or CeCoIn$_{5}$), where the heavy-fermion state appears for Ce$^{+3+\delta }$, with $0 \leq \delta \ll 1$, i.e. for the band filling of the correlated narrow $f$-band $n_{f}=1-\delta$. 
 
The question arises whether such a decomposition can occur also for $3d$ electrons.\cite{pl1,PVD} This is a legitimate question, since the $s$-$d$ model, involving a mixture of  localized and itinerant electrons is invoked \textit{ad hoc}, \cite{nagaeev} for example for the semiconducting spinels and the manganites. In the case of orbitally degenerate states of one kind, the decomposition into the localized and the itinerant parts (\textit{partial localization}) is usually  accompanied by the \textit{symmetry change}.
Obviously, the symmetry breaking may not be required if the decomposition is realized via a discontinuous phase transition. 

The \textit{partial localization} has been discussed mainly in the model situations.\cite{gz0, pl1, PVD} In this paper we would like to discuss specific physical properties related to this phenomenon not discussed in detail so far. Namely, we show first that the transformation into the localized-itinerant mixture is  often accompanied by a formation of \textit{saturated ferromagnetic} state, i.e. with one spin orientation of carriers at the Fermi level. Secondly, the spin dependence of the mass enhancement, predicted some time ago,  \cite{SpalekGopalan} and confirmed experimentally very recently, \cite{McCollam} for the case of $5f$-electron systems, is estimated numerically to see if such interesting effects are observable also  for the itinerant $3d$ magnets. Finally, we provide an exact analytic argument how the orbitally degenerate Hubbard model with the Hund's rule coupling included,  can be transformed into an effective s-d model involving the partial localization. Thus, the concept of partial localization connects the two models regarded as physically disparate.

The structure of the paper is as follows. In Sec. II we discuss the model (and its limitations), as well as introduce the auxiliary boson representation in the saddle-point (mean-field) approximation. In Sec. III we discuss the ground state properties in both para- and ferro-magnetic states, for both integer and noninteger filling $n$. In Sec. IV we derive analytically the effective $s$-$d$ ($s$-$f$) model of correlated electrons assuming that we have a partially localized state for $ 1 < n  \leq 2 $ and show a competitive nature of ferromagnetic interactions ( double exchange, kinetic exchange among the remaining itinerant electrons) and antiferromagnetic kinetic exchange. Sec. V contains a brief list of topics not discussed in detail in this paper. 
\section{\label{sec:2} The Model}
We start from the orbitally doubly degenerate version of the Hubbard model containing inequivalent, 
but spatially isotropic, hopping integrals, $t_{ijl}=t_{l}$, with the orbital index $l=1$ or $2$. This means that the starting system Hamiltonian is of the form:

\begin{eqnarray}
\mathcal{H}  &=& \sum_{\langle
i,j \rangle,\sigma} \sum_{l=1}^{2} t_{l}a^{\dag}_{il\sigma}a_{jl  \sigma}-2J\sum_{i} (\mathbf{S}_{i1} \cdot \mathbf{S}_{i2}+ \frac{3}{4}\hat{n}_{i1}\hat{n}_{i2})
\nonumber \\
&+&
 U \sum_{i,l} \hat{n}_{il\uparrow}\hat{n}_{il\downarrow} + (U-J)\sum_{i} \hat{n}_{i1}\hat{n}_{i2}- 2\sum_{i,l} S^{z}_{il}\cdot h. \nonumber \\
\label{fullstartinghamiltonian}
\end{eqnarray}
In this Hamiltonian the first term is the hopping term, the second expresses the complete form of intraatomic interorbital exchange (the Hund's rule coupling), the third describes  intraatomic intraorbital Coulomb term (the Hubbard term), the fourth describes the intraatomic interorbital Coulomb term with $n_{il}=\sum_{\sigma}n_{il\sigma}$ being the number of electrons on site $i$ and orbital $l$,  whereas the last term represents the effect of the external magnetic field. The primed summation means that $l \neq l^{\prime}$ and $\langle
i,j \rangle$ denotes the pair of nearest neighboring sites $i$ and $j$. Strictly speaking, we have neglected the intrasite pair-hopping $\frac{1}{2}J \sum_{i, l, l^{\prime}} a^{\dag}_{il\uparrow}a^{\dag}_{il\downarrow}a_{il^{\prime} \downarrow}a_{il^{\prime} \uparrow}$. Also, we have neglected the intersite interorbital- hybridization term, $\sum_{\langle i,j \rangle l, l^{\prime}} t_{ll^{\prime}}a^{\dag}_{il\sigma}a_{jl^{\prime}\sigma}$. While the last two omissions form a standard model for the \textit{orbital-selective (partial)} Mott-Hubbard localization,\cite{PVD} their inclusion would enrich the main features singled-out in the present paper.  Those additional terms should be included in a full version of the model, after the qualitative features have been discussed in the simplest situation, as provided below. It is our aim to show that even in the limit of isotropic hopping integrals their inequivalence $(t_{1} \neq t_{2})$ leads already to the interesting phenomena. Also,  the Hubbard parameter $U$ has been assumed as orbital independent, since we assume that the states are of the same ($3d$) type. The same concerns also the relation between inter- and intra orbital interactions in (\ref{fullstartinghamiltonian}).
\subsection{\label{subsec:1} Slave-boson approach}
The method we use is the auxiliary (slave) boson approach, in which each $K$-particle state on site $i$ is labeled with a boson field $\beta^{(K) \dag}_{i, l_{1}\sigma_{1}, \ldots,  l_{K}\sigma_{K}}$. In effect, the physical $K$-electron state located on that site assumes the form
\begin{eqnarray}
|i, l_{1}\sigma_{1}, \ldots, l_{K}\sigma_{K} \rangle  =  
\prod_{m=1}^{K}a^{\dag}_{il_{m}\sigma_{m}}| 0 \rangle =  \nonumber \\ 
\beta^{(K) \dag}_{i, l_{1}\sigma_{1}, \ldots,  l_{K}\sigma_{K}} \prod_{m=1}^{K}f^{\dag}_{il_{m}\sigma_{m}} | v \rangle,
\end{eqnarray}
where $| v \rangle$ is the auxiliary vacuum state and $f^{\dag}$ represents the pseudofermion creation operator. The new Fock space contains states which have no physical meaning. To get rid of them, we have to take into account the following constraints:
\begin{eqnarray}
\sum_{K=0}^{4} \sum_{I_{K}} \beta^{(K)\dag}_{i,I_{K}}\beta^{(K)}_{i,I_{K}} = 1_{i},
\label{constraintone}
\end{eqnarray}
\begin{eqnarray}
\hat{n}_{il\sigma}=f^{\dag}_{il\sigma}f_{il\sigma} =\sum_{K=1}^{4} \sum_{I_{K}}^{ ~~~~~\prime } \beta^{(K)\dag}_{i,I_{K}}\beta^{(K)}_{i,I_{K}},
\label{constrainttwo}
\end{eqnarray}
where $I_{K}= \{l_{1}\sigma_{1}, \ldots, l_{K}\sigma_{K}\}$ is a multiindex, and primed summation is taken over configurations with $(l, \sigma)$ state occupied. The first constraint ensures the completeness  condition of the basis vector set on each site, the second expresses  the equivalence of counting the electrons in terms of either fermions or bosons.
This representation has been introduced some times ago. \cite{
KotliarRuckenstein,fresardkotliar1} It has a drawback in the sense that it does not reproduce the spin-flip part of the full Hund's-rule term through the slave bosons.\cite{rotationallyinvariant, klejnbergspalek3} Thus, within this method we can include only the Ising part of that term. In other words, we start not from the full form of the Hamiltonian (\ref{fullstartinghamiltonian}), but  from its simplified form without the spin-flip term. In effect, we can rewrite (\ref{fullstartinghamiltonian}) in the form
\begin{eqnarray}
\mathcal{H} &=& \sum_{\langle
i,j \rangle,\sigma} \sum_{l=1}^{2} t_{l}a^{\dag}_{il\sigma}a_{jl  \sigma} 
+ U\sum_{i,l} \hat{n}_{il\uparrow}\hat{n}_{il\downarrow}\nonumber \\
&+&
\sum_{i,\sigma} (U_{a} \hat{n}_{i1\sigma}\hat{n}_{i2\bar{\sigma}} + U_{p}\hat{n}_{i1\sigma}\hat{n}_{i2\sigma}) + H_{Z}.
\label{startinghamiltonian}
\end{eqnarray}
Here $U_{a}=U-2J$, $U_{p}=U-3J$, $  H_{Z} = - 2\sum_{i,l} S^{z}_{il}\cdot h$. 

The Hamiltonian  (\ref{startinghamiltonian}) expressed through the new fermion and boson fields, with inclusion of the constraints (\ref{constraintone}) and (\ref{constrainttwo}) via the corresponding Lagrange multipliers and the additional renormalizing factors \cite{KotliarRuckenstein,fresardkotliar1} reads now:
\begin{eqnarray}
\tilde{\mathcal{H}} &=& \sum_{l=1}^{2} \sum_{
i,j ,\sigma} f^{\dag}_{il\sigma}(t_{l}\hat{z}^{\dag}_{il\sigma}\hat{z}_{jl\sigma} - \sigma h \delta_{ij})f_{jl  \sigma} \nonumber \\ & + &\sum_{i} \sum_{K=2}^{4} \sum_{I_{K}} \sum_{a, b} U_{\ l_{a}\sigma_{a} l_{b}\sigma_{b}} \beta^{(K)\dag}_{i,I_{K}}\beta^{(K)}_{i,I_{K}}  \nonumber \\ & + & \lambda^{(1)}_{i}(\sum_{K=0}^{4} \sum_{I_{K}} \beta^{(K)\dag}_{i,I_{K}}\beta^{(K)}_{i,I_{K}} - 1_{i})  \nonumber \\ & + &  \lambda^{(2)}_{il\sigma}(f^{\dag}_{il\sigma}f_{il\sigma} - \sum_{K=1}^{4} \sum_{I_{K}} \beta^{(K)\dag}_{i,I_{K}}\beta^{(K)}_{i,I_{K}}),
\label{slavebosonhamiltonian}
\end{eqnarray}
where the factor renormalizing the hopping term is  
\begin{equation}
\hat{z}_{il\sigma}= \frac{1}{\sqrt{(1-\hat{n}_{il\sigma})}} \sum_{K=1}^{4} \sum_{\tilde{I}_{K-1}} \beta^{\dag (K-1)}_{i\tilde{I}_{K-1}}\beta^{(K)} _{iI_{K}}  \frac{1}{\sqrt{\hat{n}_{il\sigma}}}
\end{equation}
The factors $1/\sqrt{(1-\hat{n}_{l\sigma})}$ and $1/\sqrt{\hat{n}_{l\sigma}}$ ensure the proper Hartree-Fock limit value of $\hat{z}_{l\sigma}$.\cite{KotliarRuckenstein,fresardkotliar1}
The Hamiltonian in the form (\ref{slavebosonhamiltonian}) is used next to construct the partition function expressed as a functional integral over coherent states of Fermi and Bose fields, in a standard manner.\cite{NegeleOrland} This integral, however, cannot be handled directly, as only bilinear fermionic part can be integrated out exactly. 
To proceed further, an approximation scheme must be developed. 
We shall use the saddle-point (mean-field) approximation for the Bose fields (their mean-field amplitudes are defined in Table I, and the new labeling of the electron configurations is explicitly specified).

\begin{table}
\begin{center}
\caption{ Site configurations with $K=1,\ldots, 4$ electrons and their slave-boson labelling.}

\begin{tabular}{rrr}\hline
Configuration ~&~~ SB representation~ & ~~mean-field value \\ 
   & & of the Bose field \\ 
\hline 
$|0\rangle $ &  ~~~$e^{\dag}| v \rangle$ & $e$ \\
 $|l\sigma\rangle $ & ~~~$f^{\dag}_{l\sigma} p_{l\sigma}^{\dag}| v \rangle$ & $p_{l\sigma} $ \\ 
 $|l\uparrow l\downarrow\rangle  $& ~~~$f^{\dag}_{l\uparrow}f^{\dag}_{l\downarrow}d_{l}^{\dag}| v \rangle$ & $d_{l}$\\ 
 $|1\sigma 2 \sigma\rangle  $&  ~~~$f^{\dag}_{1\sigma}f^{\dag}_{2\sigma}d_{\sigma}^{\dag}| v \rangle$ &  $d_{\sigma} $\\ 
 $|l\sigma \bar{l}\bar{\sigma} \rangle  $&  ~~~$f^{\dag}_{1\sigma}f^{\dag}_{2\bar{\sigma}}w_{\sigma}^{\dag}| v \rangle$ & $w_{\sigma}$ \\ 
 $|l\bar{\sigma} \bar{l}\sigma \bar{l}\bar{\sigma}\rangle  $ & ~~~ $f^{\dag}_{l\bar{\sigma}}f^{\dag}_{\bar{l}\sigma}f^{\dag}_{\bar{l}\bar{\sigma}}t_{l\sigma} ^{\dag}| v \rangle$ & $ t_{l\sigma} $ \\ 
 $|1\uparrow 1\downarrow 2\uparrow 2\downarrow \rangle   $ &  ~~~$f^{\dag}_{1\uparrow }f^{\dag}_{1\downarrow}f^{\dag}_{2\uparrow }f^{\dag}_{2\downarrow}q^{\dag}| v \rangle$ & $q$ \\ 
\hline
\end{tabular}
\end{center}

\end{table}
 
\subsection{\label{subsec:2} The saddle-point approximation}
In the saddle-point approximation all the Bose fields are approximated by their expectation values. This means that the operator quantities $\hat{z}_{il\sigma}$ ($\hat{z}^{\dag}_{il\sigma}$ ) reduce to the site-independent real numbers $z_{l\sigma}$, which renormalize the bare hopping integrals $t_{l}$ and make them explicitly spin-dependent. In result, in the spin-polarized state the effective masses of quasiparticles represented by the pseudo-fermion fields become also spin dependent.  
We assume also rectangular (featureless) form of (bare) density of states in both bands: 
\begin{equation}
\rho_{l}(\epsilon)= \frac{1}{W_{l}}\theta \Big(\frac{W_{l}}{2}-|\epsilon|\Big),
\end{equation}
where $\theta $ is the Heaviside step function, and $W_{l}$ is  the bare bandwidth of the $l$-th band.
In what follows we take the limit of zero temperature. Those assumptions allow us to find closed, analytic expression for the ground-state energy function (per site) of the system, which  has the following form: 

\begin{eqnarray}
E &=& - \sum_{l,\sigma} \frac{W_{l}}{2} (e p_{l\sigma} + d_{l} p_{l\bar{\sigma}} + d_{\sigma}p_{\bar{l}\sigma}  + w_{\sigma(l)} p_{\bar{l}\bar{\sigma}}+ w_{\bar{\sigma}(l)}t_{\bar{l}\bar{\sigma}} \nonumber \\ 
&+&  d_{\bar{l}}t_{l\bar{\sigma}} + d_{\bar{\sigma}} t_{\bar{l}\sigma}+ t_{l\sigma} q)^{2} + \sum_{l} U_{l} d^{2}_{l} +  U_{a}\sum_{\sigma }w^{2}_{\sigma}  \nonumber \\ &+&  U_{p}\sum_{\sigma}d^{2}_{\sigma} + 
 \sum_{l,\sigma}( U_{\bar{l}} + U_{a} + U_{p} )t^{2}_{l\sigma}  \nonumber \\ &+& (2 U_{a} + 2 U_{p} + \sum_{l}U_{\bar{l}} )q^{2}   -\sigma h \sum_{l} n_{l\sigma}. 
\label{GroundEnergy}
\end{eqnarray}
However, one has to keep in mind that variables in (\ref{GroundEnergy}) are not independent.
First, we write down the mean-field version of the constraints  (\ref{constraintone}) and (\ref{constrainttwo}), which are:  
\begin{eqnarray}
1&=&e^{2}+\sum_{l,\sigma}(p^{2}_{l\sigma}+t^{2}_{l\sigma})+\sum_{l}d^{2}_{l}+\sum_{\sigma }(w^{2}_{\sigma}+d^{2}_{\sigma}) +q^{2}, \nonumber \\ && 
\label{mfconstraint1}
\end{eqnarray}
and
\begin{equation}
n_{l\sigma}=  p^{2}_{l\sigma} + d^{2}_{l} +  d^{2}_{\sigma}  + w^{2}_{\sigma(l)} + t^{2}_{\bar{l}\sigma} + t^{2}_{l\bar{\sigma}} + t^{2}_{\bar{l}\bar{\sigma}} + q^{2},
\label{mfconstraintnlsigma}
\end{equation}
where $n_{l\sigma}$ = $\langle  f^{\dag}_{il\sigma}f_{il\sigma}\rangle$. 
The squares of the corresponding mean-field amplitudes have the interpretation of probabilities of finding the respective electron configurations. 
It is the total number of electrons per site, $n= \sum_{l\sigma} n_{l\sigma}$, which is fixed. Thus, when minimizing $E$ we use the following constraint:
\begin{eqnarray}
n &=& \sum_{l,\sigma}( p^{2}_{l\sigma} +3t^{2}_{l\sigma})+ \sum_{l}2d^{2}_{l} + \sum_{\sigma }(2w^{2}_{\sigma}+2d^{2}_{\sigma}) + 4q^{2}. \nonumber \\ && 
\label{mfconstraintn}
\end{eqnarray}
Note, that the Lagrange multipliers do not appear in (\ref{GroundEnergy}) explicitly, as they are now expressed through the bosonic fields.
The functional dependence of the energy on the values of $e , p_{l\sigma}, \ldots, q$ leads to the result similar to those obtained earlier by means of the Gutzwiller ansatz.
The quasiparticle mass enhancement is connected to the band narrowing factor $ q_{l\sigma} \equiv  z^{2}_{l\sigma}$ by the relation \cite{SpalekGopalan}
\begin{equation}
\frac{m^{\ast}_{l\sigma}}{m_{l}}= \frac{1}{q_{l\sigma}},
\end{equation}
where $m_{l}$ is the bare (band) mass in the $l$-th band, and  
\begin{eqnarray}
q_{l\sigma}&=& (e p_{l\sigma} + d_{l} p_{l\bar{\sigma}} + d_{\sigma}p_{\bar{l}\sigma}  + w_{\sigma(l)} p_{\bar{l}\bar{\sigma}}+  
 t_{\bar{l}\bar{\sigma}}w_{\bar{\sigma}(l)} + d_{\bar{l}}t_{l\bar{\sigma}}  \nonumber \\ &+&d_{\bar{\sigma}} t_{\bar{l}\sigma}+ t_{l\sigma} q)^{2}/n_{l\sigma}(1-n_{l\sigma}) \equiv 
\gamma_{l\sigma}/n_{l\sigma}(1-n_{l\sigma}), \nonumber \\
& & 
\end{eqnarray}
where the equation defines also the quantity $\gamma_{l \sigma} $, which we refer to as the reduced band narrowing factor.
Hence, the mass enhancement factor is spin-dependent in either ferromagnetic metallic (FM) state or in paramagnetic metallic (PM) state in an applied magnetic field. It is pronounced close to the PM$\rightarrow$FM phase transition, as discussed below. Obviously, the ground-state properties of the correlated Fermi liquid are determined when  $E$ is minimized with respect to all Bose fields and with self-consistently adjusted position of the chemical potential for given band filling $n$ and for the fixed values of parameters: the bandwidths $W_{l}$ and the interaction parameters $U$ and $J$, where  here  we take that $J=0.25~U$. 
Throughout the paper we assume also that $W_{1}=4$, $W_{2}=2$. 
\section{\label{sec:3} Discussion of results}
In this Section we discuss the results obtained by the numerical minimization of the ground-state energy function (\ref{GroundEnergy}).
First, we minimize $E$ with respect to all sixteen slave-field amplitudes, with the constraints (\ref{mfconstraint1}) and (\ref{mfconstraintn}) included.
This allows us to determine  both the ferromagnetic and the paramagnetic solutions. 
Because $E$ function has the obvious symmetry with respect to the  spin index reversal, we add the symmetry-breaking constraint $n_{\uparrow} -n_{\downarrow} \geq 0$ to single out one of the two  ferromagnetic minima. 
However, minimization in the full 16-dimensional parameter space is not always an easy task, especially near the paramagnetic-ferromagnetic phase transition, where those two minima become degenerate. Hence, we restrict \textit{a posteriori} number of variables by putting some of them equal to zero in accordance with the results of the full minimization procedure. This allows us to improve the numerical accuracy of obtained solutions.

Before we provide our numerical results in detail, let us pose the question about a possible evolution of the system with the increasing $U$, for fixed band filling $n$.
There are three  \textit{a priori} possible scenarios, namely: i) both bands localize simultaneously, ii) both bands localize, but for different values of  interaction magnitude $U$, iii)  the narrower band localizes, whereas the wider band remains itinerant. Below we analyze all the three situations.
\subsection{\label{ssubsec:1} Integer band filling, n=1 and 2}

\begin{figure}[h!]
\begin{center}
\rotatebox{0}{\scalebox{0.5}{\includegraphics{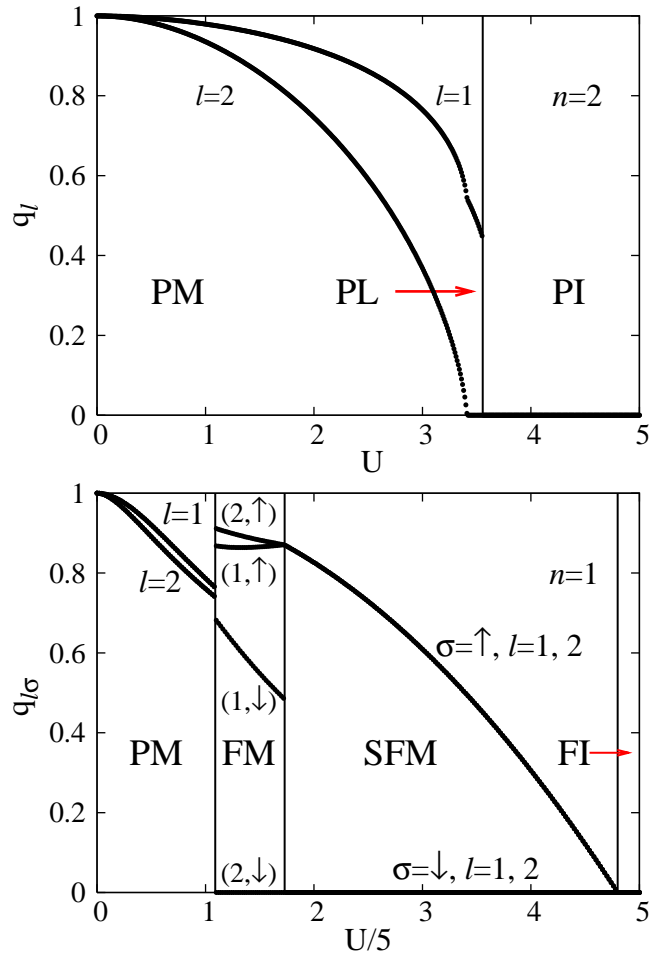}}}
\end{center}

\caption{top: $U$ dependence of the band-narrowing factors $q_{1}$ and $q_{2}$, for the half filling. The phases considered are paramagnetic metallic (PM) and insulating (PI) as well as the partially localized  (PL) phase (see the main text.) Bottom: the corresponding plot of $q_{l\sigma} $ for the quarter filling. The  phase transitions to the ferromagnetic metallic (FM), to the saturated ferromagnetic metallic (SFM) and to the ferromagnetic insulating (FI) phases are marked by the vertical lines.}
\label{PVD_and_qf}
\end{figure}

Consider first the case of half filling, $n=2$, discussed intensively recently by many authors \cite{PVD} in the context of \textit{orbitally-selective Mott transition} in Ca$_{2-x}$Sr$_{x}$RuO$_{4}$ system. 
In Fig. \ref{PVD_and_qf} we have plotted the band narrowing factor $q_{l\sigma}$, including respectively only the paramagnetic states in the half-filled case (top),  and with the ferromagnetic states included  for the quarter-filled band case (bottom). In the former case  the electrons in the narrower band can become localized in a continuous manner  at the  critical interaction magnitude $U_{c2}=3.4$, whereas those in the wider band are still itinerant and localize via first-order transition for $U_{c1}=3.56$. The intermediate region is than called \textit{partially localized} (PL) phase. Note, that all displayed phases  are of paramagnetic character (no ferromagnetic solution has been found stable). These results are very similar to those of R\"{u}egg et al.\cite{PVD}, but differ from those of van Dongen et al.\cite{PVD} obtained within QMC-DMFT method, where the character of the phase transitions is different. 

Next, we examine the quarter-filled case, i.e. $n=1$ (cf Fig. \ref{PVD_and_qf}, bottom), showing  the band narrowing factor $q_{l\sigma}$.
For this particular situation, the system transforms with the increasing $U$ first discontinuously into a ferromagnetic metallic (FM) state, with the electrons in the narrower band being fully polarized. The wider band is then partially 
polarized, unlike in the equivalent-band model, with the orbital ordering included.\cite{klejnbergspalek3} By increasing $U$ further we observe a disappearance of the minority-spin electrons in the wider band, as the system undergoes a transition to the saturated ferromagnetic metallic (SFM) phase. For sufficiently  high $U$, electrons in both bands localize simultaneously, forming ferromagnetic insulating (FI) state.
One should note, that SFM-FI phase boundary located at $U=24$, is quite analogous to the original \textit{Brinkman-Rice quantum critical point} occurring in the nondegenerate band, except, that now the transition takes place between the ferromagnetic states. In the present situation the orbital index $l=1,2$ replaces the spin quantum number for the nondegenerate case. Explicitly, only the variables  $e=d_{\uparrow} \equiv x,~~p_{1\uparrow} $, and $ p_{2\uparrow}$ have non-zero values in SFM state. Thus, we can write down the ground-state energy function $E$ (per site) of SFM state in the following analytic form:
\begin{equation}
E=-Wx^{2}(1-2x^{2})+ (U-3J)x^{2},
\end{equation}
where  $W=(W_{1}+W_{2})/2$. This expression is formally identical with that for the single-band case.  \cite{encyclopedia} Minimizing $E$ with respect to $x^{2}$, we obtain the physical ground-state characteristics:

\begin{equation}
E = E_{G}=-\frac{W}{4}\Big( 1- \frac{(U-3J)}{2W}\Big)^{2},
\end{equation}
and 
\begin{equation}
q_{l\uparrow}= 1- \Big(\frac{U-3J}{2W}\Big)^{2}.
\label{q_od_UW}
\end{equation}
 Eq. (\ref{q_od_UW}) provides the justification for the corresponding parabolic dependence of $q_{l \uparrow}$ in the SFM state, as shown in Fig.~1 (bottom).
\subsection{\label{ssubsec:2} Partially filled case, n=1.1 and 1.9}
\begin{figure}[h!]
\begin{center}					 
\rotatebox{0}{\scalebox{0.5}{\includegraphics{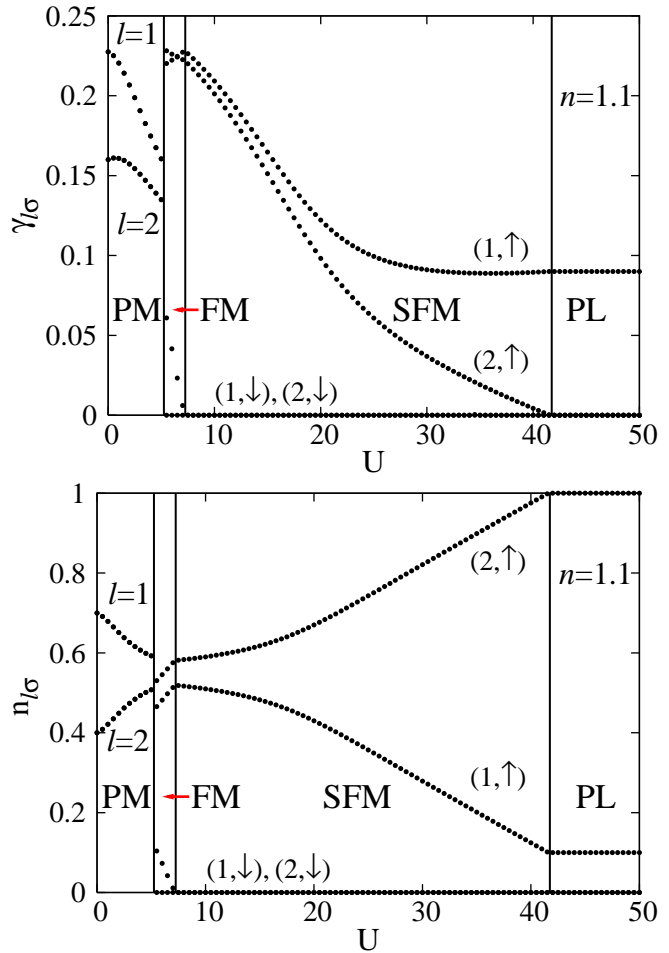}}}
\end{center}
\caption{top: $U$-dependence of the reduced band narrowing factor $\gamma_{l \sigma} = q_{l \sigma} n_{l \sigma}(1 - n_{l \sigma}) $. Bottom: the band occupancies versus U, for $n=1.1$. The phases and the corresponding boundaries are separated by the vertical lines. }
\label{q11002}
\end{figure}

For the band filling slightly larger then $1.0$, e.g. for $n=1.1$, the simultaneous localization in both bands is no longer possible, as now obviously some double occupancies must be present for an arbitrarily high value of $U$ to fulfil the constraint (\ref{mfconstraintn}). Guided by the experience gained from the $n=1$ case, we can make a conjecture about the character of the ground state in the present situation. Thus, with the increasing $U$ we expect, that the system should undergo a transition from paramagnetic metallic (PM) to a ferromagnetic state. This ferromagnetic state, in turn,  should become saturated (SFM state) for high enough $U$ resulting in state  of the same kind, as that in the ($n=1$) case. That is, only the variables $e$, $d_{\uparrow}$, $p_{1\uparrow} $, and $ p_{2\uparrow}$ have non-zero values in this state.  However, as now $d^2_{\uparrow} -e^2=0.1 \neq 0$, we expect that for sufficiently  high $U=U_{c}$ we find $e=0$, in order to minimize Coulomb repulsion, and then $E$ is minimized for $p_{1\uparrow}=0$. Thus, for $U > U_{c}$ electrons in the narrower band localize, resulting a in simple PL phase, for which all the variables except $d_{\uparrow}$ and $p_{2\uparrow}$ have nonzero values. In other words, electrons in the  narrow band are localized and that band is fully occupied, forming a spin background for itinerant ($1 \uparrow$) quasiparticles. 
Those predictions are confirmed by the detailed numerical analysis. In Fig. \ref{q11002}, top, we have plotted the reduced band narrowing factor $\gamma_{l\sigma}$ instead of $q_{l\sigma}$, as its value  is well-defined for values of $n_{l\sigma}$ close to zero or unity. In the bottom part of Fig. \ref{q11002} we have plotted the corresponding occupation numbers. Interestingly enough, we see that, firstly the FM phase exists in a relatively narrow interval of $U$, and secondly, that the SFM-PL transition occurs for rather high value of $U_c \approx 41.5 $.    

The results obtained for $n=1.0$ and $n=1.1$ suggest that for large $U$ the ground state of the system is ferromagnetic.
However, with the increasing band filling the character of this state changes  as for $n \to 2$ it is no longer favorable as a nonzero value of $e$ should be retained.

\begin{figure}[h!]
\begin{center}
\rotatebox{270}{\scalebox{0.35}{\includegraphics{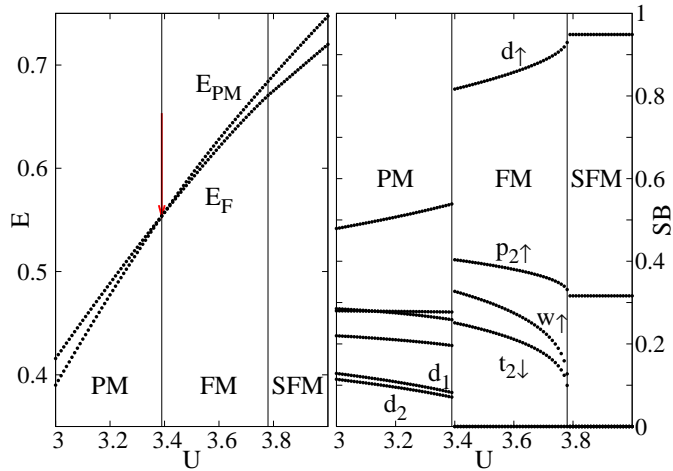}}}
\end{center}

\caption{right: $U$ dependence of the ground-state energies of paramagnetic (PM) and ferromagnetic (F) solutions. Left: values of selected slave-boson mean field amplitudes. Phase transitions are marked by the vertical lines.}
\label{multiplotenergieisb19}
\end{figure}

\begin{figure}[h!]
\begin{center}
\rotatebox{270}{\scalebox{0.35}{\includegraphics{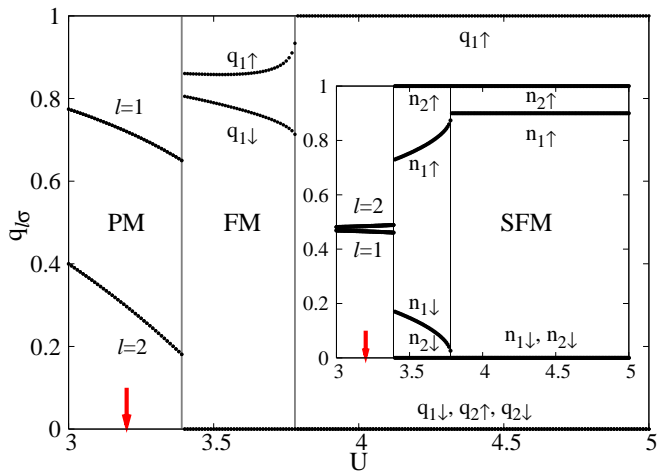}}}
\end{center}

\caption{$U$ dependence of the band narrowing factors $q_{l \sigma}$ for n=1.9. Inset: corresponding occupation numbers. Phase transitions are marked by the vertical lines.}
\label{multiplotnq19}
\end{figure}

\begin{figure}[h!]
\begin{center}
\rotatebox{0}{\scalebox{0.5}{\includegraphics{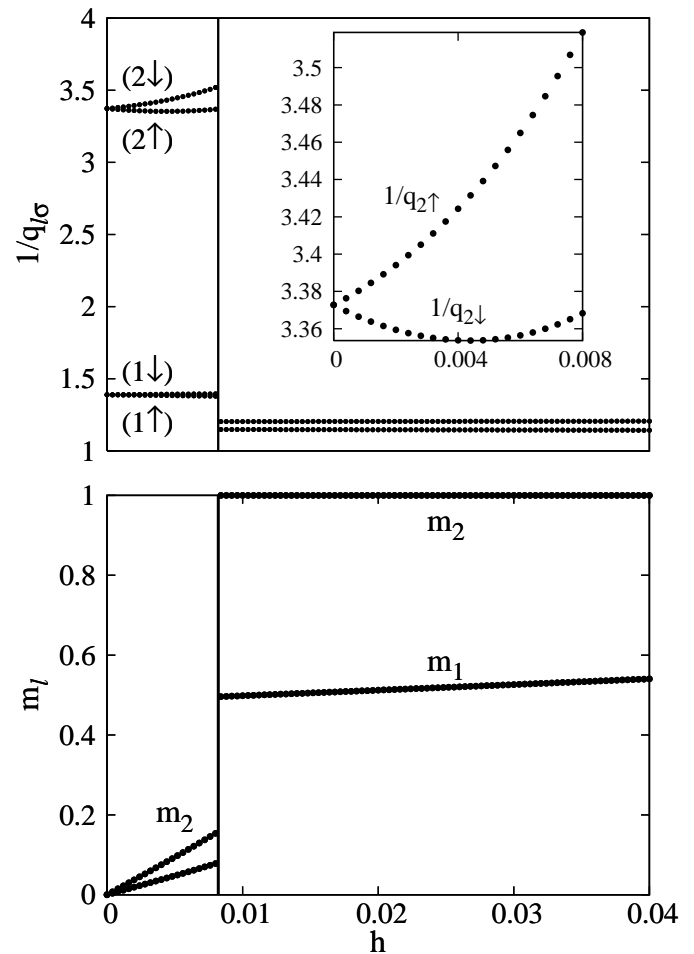}}}
\label{magimasy}
\end{center}

\caption{top: applied field-dependent mass enhancements for  $n=1.9$ and $U=3.2$; 
bottom: corresponding orbital dependent spin polarizations ($m_{l}= n_{l \uparrow} - n_{l \downarrow}) $ for both bands. Phase transition is marked by the vertical line.
Inset: detailed mass enhancement for the narrower band below the localization threshold.}
\label{masses_and_mag}
\end{figure}

\begin{figure}[h!]
\begin{center}
\rotatebox{270}{\scalebox{0.35}{\includegraphics{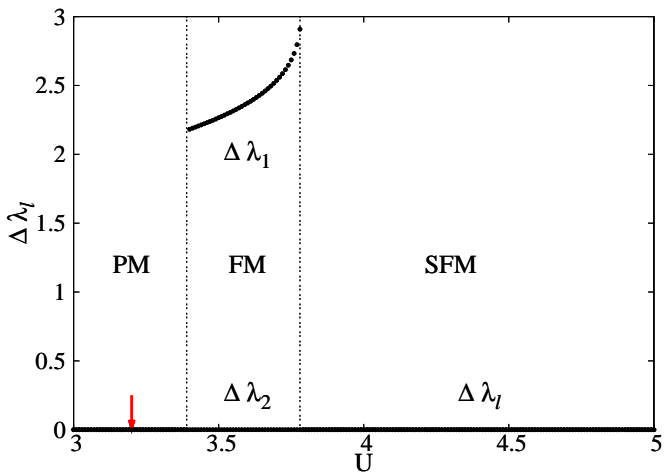}}}
\end{center}

\caption{ Plot of $\Delta \lambda_{1} \equiv (\lambda_{1\downarrow}- \lambda_{1\uparrow})/J$, $\Delta \lambda_{2} \equiv (\lambda_{2\downarrow}- \lambda_{2\uparrow})/J$ vs. $U$ for $n = 1.9$}
\label{roznicapotchem}
\end{figure}

Next, we analyze the  situation for the band filling $n=1.9$, (c.f. Figs. \ref{multiplotenergieisb19}-\ref{roznicapotchem}). In Fig. \ref{multiplotenergieisb19}, left, we display the energies of paramagnetic (PM) and ferromagnetic (F) solutions, respectively, and show that for $U= 3.4$ we have the paramagnetic - ferromagnetic transition.  To reveal the nature of those states, let us analyze first the right part of Fig. \ref{multiplotenergieisb19}, where some of the  slave boson amplitudes are plotted. For $U < 3.4$ we have an ordinary paramagnetic behavior, very much alike in the earlier cases. 
However,  for $3.4 < U < 3.78$  only the variables $ p_{2\uparrow},  d_{\uparrow},  w_{\uparrow}$, and $ t_{2\downarrow}$ have nonzero values. In other words, we have always one $(2\uparrow)$  electron (forming the spin-background) and one of the four possible configurations for the $l=1$ orbital. Thus, in this case we also have effectively a one-band behavior, for which in the limit $U > 3.78 $ only $ p_{2\uparrow}$ and $  d_{\uparrow}$ are nonzero, just like for the corresponding PL phase for $n=1.1$ case. In this state both $d_{1}=d_{2}=0$, even though the system is only partially spin-polarized. This situation is 
different from that for a non-degenerate-band case, since now we can still have substantial interorbital spin-singlet correlations (see also below).

In Fig. \ref{multiplotnq19}, we display both the band narrowing factors $q_{l\sigma}$ and the occupancies (cf.~ the inset) In the paramagnetic (PM) phase electrons in both bands retain itinerant character. The PM-F transition represents now also the localization threshold for electrons in the narrower band which becomes completely localized, polarized and form a spin background with $n_{2\uparrow}=1$, $n_{2\downarrow}=0$ (see  the inset).
Additionally, both the $(1\uparrow)$ and  $(1\downarrow)$ electrons remain itinerant for $3.4 < U < 3.78$ composing together with $(2\uparrow)$ electrons a partially localized ferromagnetic metallic phase (FM),  which subsequently transforms into SFM state, where only $(1\uparrow)$ electrons remain itinerant.  One should note that the residual $(n-1)$ carriers per site acquire the bare band mass in this state ($q_{1\uparrow}=1$). This is because the Hubbard interaction vanishes in the SFM state and the hopping is not hampered by the Hund's rule interaction.

The PM-FM transition is realized with the increasing amplitude of $U$. There is however, another possibility, namely, we can induce it by applying an external magnetic field in the paramagnetic state as shown in Fig.5. Apart from this transition we expect also that an effective mass of electrons will become then also spin-dependent.
To analyze these effects  in detail, we have drawn in Fig. \ref{masses_and_mag} (top) their enhancement as a function of $h = g\mu_{B}H_{a}/2$ for the partially filled-band configuration $n=1.9$. The masses in the narrower ($l=2$) band become infinite (i.e. electrons localize) when the system undergoes a \textit{metamagnetic transition} (as shown in the bottom part of the figure). If the model parameter values specified there, are taken in electronovolts, then the metamagnetic field is of the order of $100 $ T and should diminish fast  with $n \rightarrow 2$. The ferromagnetic phase is stable even for $n=1.9$, when the Hund's rule is strong enough and overcomes the tendency towards the antiferromagnetism (not discussed here.) 
 In the inset to Fig. \ref{masses_and_mag} (top panel) we display  the masses in the low-field range, and show their nonlinear field dependence. The important feature of the transition in the applied field is that it requires relatively low external perturbation in the range of meV, whereas the corresponding critical interaction $U$ needed for the transitions displayed in Fig.~2 is in the eV range. However, near the Mott localization the renormalized band and the interaction parts of the total energy are reduced and almost compensate each other. Hence, this relatively small perturbation shifts the balance between them in favor of partially localized state.

In Fig. \ref{roznicapotchem} we have determined the effective exchange field acting on the correlated electrons. It originates from the explicitly spin-dependent constraint (\ref{constrainttwo}), and is related to the Lagrange multiplier $\lambda^{(2)}_{l \sigma}$, defined in (\ref{slavebosonhamiltonian}). Their difference is displayed in Fig. \ref{roznicapotchem} in units of $J$ and  as  function of $U$. One should note that the exchange field difference $\Delta \lambda_{l} \equiv (\lambda^{(2)}_{l\downarrow}-\lambda^{(2)}_{l\uparrow})/J$ is of the order of $U$. It is this effective field, which in conjunction with the Hund's rule interaction stabilizes ferromagnetism in a wide range of the band filling $n <2$.

\begin{figure}[h!]
\begin{center}					 
\rotatebox{270}{\scalebox{0.35}{\includegraphics{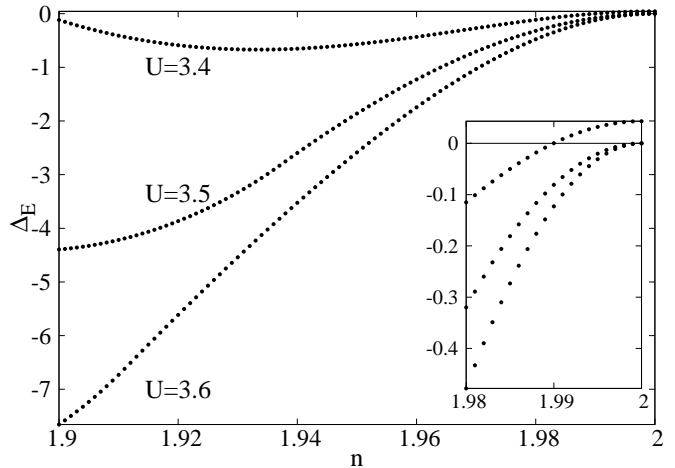}}}
\end{center}
\caption{Band-filling dependence of the FM-PM energy difference, where $E_{F}-E_{P} =\Delta_{E}/10^3 $  for the interaction values $U=3.4$, $U=3.5$, and  $U=3.6$, respectively.  Inset: detailed behavior in the region near half-filling. }
\label{growing_n}
\end{figure}

Finally, the qualitative difference between the $n=1.9$ and the half-filled cases (c.f. Fig \ref{PVD_and_qf} top and Fig. \ref{multiplotnq19}) poses the question about the nature of the ground state in the range $1.9 \leq n \leq 2.0$. To address this we show in Fig. \ref{growing_n}  the ground-state energy dependence  on $n$ in this interval for three fixed values of Coulomb interaction. For $U=U_{c1}=3.4$ there is a finite filling interval $1.99 < n < 2.00$ in which paramagnetic state is more stable than ferromagnetic one. For larger values of Coulomb interaction, $U=3.5$ and $3.6$, however, ferromagnetic state becomes stable for any non-half filling (e.g. up to $1.999$), whereas exactly at half filling  those two states become energetically degenerate. However, since the difference is  small,  the antiferromagnetic solution can easily become the stable state then.  However, one should underline that appearance of stable ferromagnetic state in the wide range of the filling is facilitated by the Hund's rule exchange.

\section{\label{sec:4} Effective s-d (s-f) model}

From the discussion above a clear division into the localized and itinerant electrons emerges for $ 1 < n < 2$, when the interaction is strong enough.  This division is achieved via a phase transition in which $\sum_{l\sigma}n_{il\sigma}$ electrons per site decomposes into p localized and $n_{i\sigma}$ itinerant particles. We provide now a simple analytic argument and show that in the partially localized situation the model represented by Hamiltonian (\ref{fullstartinghamiltonian}) reduces to an effective $s$-$d$ ($s$-$f$) model with a proper form of kinetic exchange interactions. The argument is valid for an arbitrary $p$, but for the present model with the orbital degeneracy ($d=2$) $p = 1$. 

 In order to deal with the interaction terms, in the situation with the localized-itinerant mixture, we use the following identities involving the interaction term in (\ref{fullstartinghamiltonian}):

\begin{equation}
 \sum_{l} \hat{n}_{il\uparrow}\hat{n}_{il\downarrow}=\frac{1}{2}\sum_{l\sigma}\hat{n}_{il\sigma} - \frac{2}{3} \sum_{l} \mathbf{S}_{il\uparrow}^2,
\end{equation}

\begin{equation}
 \sum_{l \neq l^{\prime};\sigma \sigma^{\prime}}  \hat{n}_{il\sigma}\hat{n}_{il^{\prime}\sigma^{\prime}}=(\sum_{l\sigma}\hat{n}_{il})^2 - 2 \sum_{l} \hat{n}_{il\uparrow}\hat{n}_{il\downarrow} - \sum_{l\sigma}\hat{n}_{il\sigma},
\end{equation}

and

\begin{equation}
 \sum_{l \neq l^{\prime}} \mathbf{S}_{il} \cdot \mathbf{S}_{il^{\prime}}= (\sum_{l}\mathbf{S}_{il\uparrow})^2 - \sum_{l} \mathbf{S}_{il\uparrow}^2.
\end{equation}
Next, we make the decomposition $\sum_{l} \mathbf{S}_{il}= \mathbf{S}_{i}+\mathbf{s}_{i}$,
with $\mathbf{S}_{i}^{2}=\frac{p}{2}(\frac{p}{2}+1)$.  This means, that we have subdivided the total spin into the conserved (localized) and the itinerant parts. Employing this rule, taking into account also that $\sum_{l}n_{il\sigma}= p + n_{i\sigma}$, and projecting out the double occupancies in the localized Mott state, one obtains up to a constant:

\begin{eqnarray}
{\mathcal{H}} &=& \sum_{ij\sigma} t_{ij1}a^{\dag}_{i\sigma}a_{j  \sigma} 
-2J \sum_{i} \mathbf{S}_{i}\cdot\mathbf{s}_{i}  \nonumber\\ &+&\sum_{i}U\hat{n}_{i1\uparrow}\hat{n}_{i1\downarrow} + {\mathcal{H}}_{ex}.
\label{sdhamiltonian}
\end{eqnarray}
where $t_{ij1}$ is the larger of the two hopping integrals. The hopping in the narrower band vanishes at the localization threshold because we have then $n_{i2\uparrow}+n_{i2\downarrow}=1$ and therefore, $a^{\dag}_{i2\sigma}a_{j2\sigma}=a^{\dag}_{i2\sigma}(1-\hat{n}_{i2\bar{\sigma}})a_{j2\sigma}(1-\hat{n}_{j2\bar{\sigma}})=0.$ Also, ${\mathcal{H}}_{ex}$ contains the kinetic exchange interaction in the localized states (band $2$). It has the following form \cite{habszefa} 
\begin{equation}
{\mathcal{H}}_{ex}= \sum_{\langle i,j \rangle} \Big(\frac{t^{2}_{ij2}}{U} + \frac{t^{2}_{ij2}}{U+J} \Big) \Big( \mathbf{S}_{i}\cdot\mathbf{S}_{j}-\frac{1}{4} \Big).   
\end{equation}
Obviously, this simple form of the effective Hamiltonian is still orbital-dependent, since only part of the electrons ( those in a narrower band) localize. It has the form of $s$-$d$ ($s$-$f$) Hamiltonian with the Hubbard interaction among the remaining itinerant electrons. Actually, in the case $n = 1.9$, shown in the Fig. 9, the double occupancy $d^{2}_{1} \rightarrow 0$ when electrons in the narrower band localize. In that situation, the itinerant electrons become also strongly correlated, i.e. represented by the effective Hamiltonian with the projected out double occupancies also for the itinerant states, namely:
\begin{eqnarray}
{\mathcal{H}}&=&t_{1}\sum_{\langle ij \rangle \sigma} a^{\dag}_{i1\sigma}(1 - \hat{n}_{i1\bar{\sigma}})a_{j1\sigma}(1 - \hat{n}_{j1\bar{\sigma}}) \\ \nonumber &-& 
2J \sum_{i} \mathbf{S}_{i}\cdot\mathbf{s}_{i}+{\mathcal{H}}^{ \prime}_{ex},
\label{correlatedsdhamiltonian}
\end{eqnarray}
where
\begin{eqnarray}
{\mathcal{H}}^{ \prime}_{ex}&=&\sum_{\langle i,j \rangle}\frac{t^{2}_{2}}{U}  \Big( \mathbf{S}_{i}\cdot\mathbf{S}_{j}-\frac{1}{4} \Big)\\ \nonumber 
&+& 2\sum_{\langle i,j \rangle} \Big(\frac{t^{2}_{1}}{U} + \frac{t^{2}_{1}}{U+J} \Big) \Big( \mathbf{s}_{i}\cdot\mathbf{s}_{j}-\frac{1}{4} \hat{n}_{i1}\hat{n}_{j1} \Big) \\ \nonumber 
&-& 2\sum_{\langle i,j \rangle }  \frac{t^{2}_{1}}{U-J}  \Big( \mathbf{s}_{i}\cdot\mathbf{s}_{j}+\frac{3}{4} \hat{n}_{i1}\hat{n}_{j1} \Big).
\end{eqnarray}
One should note that in this regime the natural limit is to be $|t_{2}| \approx J$, in which the $s$-$d$ exchange may become comparable to the kinetic exchange, since the strong double exchange interaction sets in. This may lead to ferromagnetism well beyond $n=1$ situation, but this topic should be analyzed separately.
\section{\label{sec:5} Conclusions}
The mixed (localized+itinerant) nature of the correlated quantum-mechanically indistinguishable electrons has been discussed for the case of a doubly degenerate narrow band.  It is connected  with the assumed difference in bandwidth, $W_{1} \neq W_{2}$. The breakdown of the particle indistinguishability is accomplished through a phase transition.
This  indistinguishability breakdown takes the extreme form in the limit of partial localization (PL), where $m_{2 \uparrow}=m_{2 \downarrow}= \infty$ and $m_{1 \uparrow} \neq m_{1 \downarrow} < \infty $. The two transitions shown in Fig. 1 (top panel) for the case of the half-filled band can be tested by applying the pressure in the case of appropriate Mott insulating system. Also, the properties obtained here should be tested further  on a  model involving realistic orbitals, including  the orbital ordering, as well as the antiferromagnetism. The most stringent test for the existence of PL state should come from the situation involving the hybridized orbitals.  The inclusion of the hybridization (i.e. with the hopping $t^{ll^{\prime}}_{ij}$ with $l \neq l^{\prime}$ would allow us to study the intermediate situation between the heavy fermion limit ($t_{2}=0, t^{12} \neq 0$, atomic degeneracy lifted) and the present situation (with $t_{2} \neq 0, t^{12} = 0$). Also, the situation in an anisotropic system modelled by e.g. doubly degenerate band of $e_{g}$ type should be considered in detail. We should see a progress along this lines in the near future.\\

\begin{acknowledgments}
The two of authors (J. J. and J. S.) were partially supported by the Ministry of Education and Science, Grant No. 1P03B 001 29 and by the Foundation for Polish Science (FNP). One of the authors (J.J) was supported through the Socrates (Erasmus) Program of the EU. The technical help and many valuable discussions with Andrzej Kapanowski and Sebastian Sapeta are also acknowledged.
\end{acknowledgments}

\end{document}